\documentclass[twocolumn,english,pra,showpacs,floatfix]{revtex4}
\usepackage[T1]{fontenc}
\usepackage[latin1]{inputenc}
\usepackage{graphicx}
\usepackage{amssymb}

\makeatletter

\usepackage{graphicx}

\usepackage{babel}
\makeatother

\newcommand \bea{\begin{eqnarray}}
\newcommand \eea{\end{eqnarray}}
\newcommand \ga{\raisebox{-.5ex}{$\stackrel{>}{\sim}$}}
\newcommand \la{\raisebox{-.5ex}{$\stackrel{<}{\sim}$}}
\newcommand{\av}[1]{\langle{#1}\rangle}
\newcommand{\e}{\bar{\epsilon}_{\bf q}}
\newcommand{\ek}{\bar{\epsilon}_{\bf k}}

\begin{document}
%\twocolumn[\hsize\textwidth\columnwidth\hsize
%\csname@twocolumnfalse%
%\endcsname
\draft
\title{Phases in optical lattices vs. Coulomb frustrated HTc cuprates}
\author{H. Heiselberg}
\address{Univ. of S. Denmark, Campusvej 55, DK-5230 Odense M, Denmark}
\email{heiselberg@mil.dk}
%\address{Danish Defense Research Establishment, Ryvangsalle' 1, 
%DK-2100 Copenhagen \O, Denmark}
\begin{abstract}
Fermionic atoms in 2D optical lattices and electrons in HTc cuprates
may both be described by the Hubbard model. However, if Coulomb
frustration is responsible for the striped phases in 2D cuprates the
phase diagrams will differ markedly. Two representative scenarios are
described by a simple stripe model without phase separation and a mean
field model with phase separation in the absence of Coulomb
frustration. When Coulomb frustrated both models display
antiferromagnetism (AF) and stripe phases with d-wave superfluidity,
whereas neutral atoms in optical lattices will only do so in the
stripe model. Radii and densities of the various phases in
harmonically confined optical lattices are calculated for the two
models and have very different Mott plateaus and density
discontinuities. Observation of antiferromagnetic, stripe
and superfluid phases in density and momentum distributions and
correlations from time-of-flight experiments is discussed.
\\
\end{abstract}
\vskip1pc]
\maketitle

\section{Introduction}

Ultra-cold atomic Fermi-gases present a new opportunity to study
strongly correlated quantum many-particle systems. Optical lattices
provide a periodical lattice potential in which the atoms are
described by the tight-binding approximation, which enables access to the
Hubbard model with interactions tuned by Feshbach resonances.
Most importantly we gain direct observation into many
controversal issues in strongly correlated high temperature (HTc)
superconductors in a controllable way where interactions, densities,
temperatures, etc., can be tuned.  Recent experiments with optical
lattices have measured momentum distributions and correlations, and
have found superfluid phases of Bose and Fermi atoms \cite{Stoferle,Folling,Ketterle}, 
Mott insulators \cite{Spielman}, and band insulators \cite{Kohl,Rom}.

HTc cuprates, however, suffer long range Coulomb repulsion between
doped electrons or holes (if localized) whereas neutral atoms in optical
lattices are not likewise frustrated. Coulomb frustration
inhibits phase separation and may also be responsible for stripe formation
in cuprates which again affects d-wave superconductivity (dSC). By
studying the phases in optical lattices we can directly observe
whether phase separation, stripes and dSC occur in the Hubbard model
without Coulomb frustration for general density, temperature, interaction
strengths, etc.

Many models have been applied for investigating HTc such as the 2D Hubbard
and t-J Hamiltonians in different approximate versions with very
different results (see \cite{Aima} and references therein).
Extending with next-nearest neighbor hopping and
interactions or long-range Coulomb interactions further complicates
the problem.
Early Hubbard mean field (MF) calculations predicted phase separation
between an antiferromagnetic (AF) phase with density $n=1$ and a paramagnetic phase for
$U\la 7t$ or a ferromagnetic phase for $U\ga7t$ \cite{Langmann}. 
Phase separation
(PS) in the Hubbard model and the closely related t-J model is still
controversial despite intense investigations. In the t-J model PS is
found for large $J/t$ but for the more realistic smaller values of
$J/t$ the field is divided. In the Hubbard model without next-nearest
neighbor interactions $(t'=0)$ exact diagonalization \cite{Moreo} and
Monte Carlo results \cite{Becca} find PS whereas dynamical mean field
\cite{Zitzler} and variational cluster perturbation theory
\cite{Aichhorn} do not. In dynamical cluster approximations PS is
found for finite $t'$ \cite{Macridin}.

A MF ground state with stripes was found by Zaanen and Gunnarson
\cite{Zaanen}. These stripes have a hole density of one per site
whereas density matrix renormalization group calculations
\cite{Scalapino} find a stripe density of 1/2 in accordance with
experiments \cite{Tranquada}. However, Green's
function Monte Carlo calculations \cite{Sorella} find only weak signs
of stripes. Various cellular dynamical MF \cite{Capone} and 
cluster calculations \cite{CC} find competing
antiferromagnetic and dSC phases.  The long range Coulomb
forces in cuprates suppress phase separation, possibly leading to mixed phases with
short range density fluctuations such as stripes \cite{Kivelson}.  
The atoms in optical lattices are electrically neutral
and the two phases of different densities are therefore not forced to
mix by Coulomb forces to e.g. a stripe phase 
but can remain in separate bulk phases.

The many different results indicate that the ground state is delicately
balanced between nearly degenerate phases. Slight approximations or
changes in hopping and interaction parameters can change the ground
state phase dramatically. Thus, two decades after the discovery of HTc,
the ground state and the mechanism of pairing remains undetermined in
cuprates as well as Hubbard and t-J models.

The purpose of this work is rather to study qualitative differences
in confined optical lattices from simple calculations for the 2D
Hubbard model in which Coulomb frustration play a central or no role
for inhibiting phase separation, stripe formation and HTc.  To this
purpose we specifically investigate in section II two simple
representative models with and without PS, their
different density distributions and phases. 
We investigate the effects of Coulomb frustration in
section III and show that it is difficult to distinguish the two
models in HTc materials because phase separation is inhibited and
stripe phases appear in both cases. Likewise dSC appears in the stripe
phases as discussed in section IV. Yet in confined optical lattices as
described in section V, the absence of Coulomb frustration leads to
very different phases, density and momentum distributions and correlations
for the two models.

\section{Equation of states for the 2D Hubbard model}

In the tight binding approximation spin 1/2 fermions are described by the 
Hubbard model \cite{Hubbard} 
\bea
 H= -t\sum_{\av{ij}\sigma} \hat{a}_{i\sigma}^\dagger \hat{a}_{j\sigma} \,+\,
  U\sum_i \hat{n}_{i\uparrow} \hat{n}_{i\downarrow} \,,
\eea
where $\hat{a}_{i\sigma}$ is the usual Fermi 
creation operator ($\sigma=\uparrow,\downarrow$), $n_{i\sigma}=\hat{a}_{i\sigma}^\dagger \hat{a}_{i\sigma}$ 
the density, and $\av{ij}$ denotes nearest neighbors. Only 2D square lattices 
at zero temperature will
be studied here.
$U$ is the on-site usually repulsive interaction and $t$ the nearest-neighbor hopping parameter. The model can be extended to include next-nearest-neighbor hopping ($t'$) \cite{Micnas,Langmann}
and longer-range hopping and interactions. We will mainly discuss the 
$t'=0$ case where the phase diagram is symmetric around half filling. It is then
sufficient to discuss only the doped case where $x=1-n\ge 0$. 
 
As mentioned in the introduction, various approximations and numerical
results to the Hubbard model lead to a variety of phases as ground
state solutions. Long-range Coulomb repulsion present for the case of
electrically charged electrons/holes further complicates the phase structure
and will be included in a later section.

In the following two subsections two simple representative models,
a MF and a stripe model, are
investigated with and without PS respectively.

\subsection{Phase Separation in Mean Field models}

MF theory provides a first impression of the phases competing for the
ground state and has the advantage that it is computationally simple
as compared to more complicated theories.  The MF equations for the
Hubbard model are standard and we refer to, e.g., Refs. \cite{Micnas,Langmann}. The
energy densities can be calculated within the Hartree-Fock
approximation for the paramagnetic (PM), ferromagnetic (FM), antiferromagnetic
(AF) and other phases. In Fig. 1 we show
an illustrative example for the energy density vs. filling fraction
for $U=6t$ and in the symmetric case $t'=0$.

At low density $n\ll 1$ the ground state is that of a dilute paramagnetic (PM) gas
with energy
\bea \label{PM}
  \varepsilon_{PM} = -4tn+\left[\frac{\pi}{2}t+\frac{1}{4}U\right]n^2 
 + {\cal O}(n^3)\,.
\eea
In units where the lattice spacing is unity ($a=1$) this energy per site is 
also the energy density, and the density is the site filling fraction.

Near half filling $\varepsilon_{PM} = -(4/\pi)^2t+U/4$, which becomes
positive when the repulsive interaction exceeds $U/t\ge 64/\pi^2\simeq
6.5$. The PM phase is then no longer the ground state.  In a state with
only one spin the antisymmetry of the wavefunction automatically
removes double occupancy and the repulsive term $Un^2/4$ in Eq. (\ref{PM})
disappears. Such a ferromagnetic (FM) state always has negative energy
$\varepsilon_{FM}\le0$ for $n\le1$ and is a candidate for the
ground state. AF, linear AF \cite{Kusko} and stripe phases are other 
competing candidates.

For the 2D Hubbard model it is,
however, necessary to extend MF to mixed phase scenarios.  When $U\la
7t$ the ground state of the MF Hubbard model undergoes transitions
from an AF at half filling to a mixed AF+PM phase for $|x|$ up to a finite
($U$ dependent) value whereafter a pure PM phase takes over.
\cite{Langmann} For larger $U$ the phase diagram is more complicated
with a pure as well as mixed FM phases between the AF and PM phases. A
finite next neighbor hopping term $t'$ makes the phase diagram
asymmetric around $x=0$, extends the AF phase and changes the phase diagram 
considerably.

At half filling $n=1$ the ground state is an AF.
Near half filling $0\le x \ll1$
the MF equations and AF energy can be expanded as
\bea
   \varepsilon_{AF} = -J[1+\frac{3}{2}x+2x^2+...] \,. 
\eea
The concave dependence on $x$ signals phase separation into a mixed
phase of AF and PM by the Maxwell construction as shown in Fig. 1. The
PS extends from the AF phase with density $n=1$ to a PM phase with $U$
dependent density $n_s=1-x_s\simeq 0.6$. The double
tangent determines the PS energy $\varepsilon_{PS}$ within MF.

\begin{figure}
\includegraphics[scale=0.5,angle=-90]{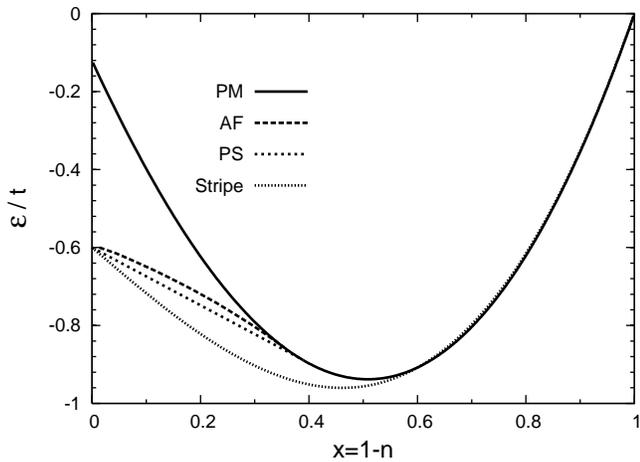}
\caption{2D Hubbard MF energy densities in the case $U=6t$ are shown
for the PM, AF and the Maxwell construction for the 
phase separation (PS). Also a stripe energy 
is shown (see text).
\label{fig1  }}
\end{figure}

The difference in energy between the PS and the PM/FM phases sets the
energy and temperature scale for the AF correlations, which is
related to 
the pseudogap, which in turn is related to the spin gap as \cite{Fradkin}
$T^*\sim\Delta_{s}/2$. 
We expect that the spin gap is of order the energy difference between
the AF phase and spin uncorrelated PM or FM phase, whichever has lowest energy.
Thus $T^*\sim (\varepsilon_{PS}-\varepsilon_{PM})/2\sim (1-x/x_s)J/2$.
The AF spin correlations have a long but finite length scale so that
the Mermin-Wagner theorem is not violated.

\subsection{Stripe model}

Stripes have been discovered in low-energy magnetic neutron scattering
in doped cuprates at incommensurate longitudinal and horizontal charge
and spin wave numbers \cite{Tranquada}, e.g. ${\bf Q}_c=(2\pi/a)(0,\pm
x/x_s)$ and ${\bf Q}_s=(\pi/a)(1,1\pm x/x_s)$ for the longitudinal charge
and spin wave numbers respectively. Here, the hole filling in the
stripes is half filled $x_s=1/2$ for low dopings $x\le 1/8$ but at
larger doping it increases linearly to filled stripes $x_s=1$ such
that $x/x_s=1/4$ remains constant.  Therefore the charge (spin)
density stripes appear with periodic distance $d$ ($2d$) depending on
doping as $d=ax_s/x$.  The stripe distance decreases with increasing
doping until $x\ge1/8$, whereafter the stripes remain at a distance
$d=4a$. The stripes act as anti-phase domain walls and the spin
density wave therefore has periodicity twice the length of the charge-density wave.

Diagonal, horizontal/vertical, checkerboard stripe solutions have been
found in a number of models. In MF models Zaanen and Gunnarson
\cite{Zaanen} found stripes of hole density $x_s=1$ that are vertical
or horizontal for $U/t\la 3-4$ and diagonal otherwise.  In DMRG
calculations \cite{Scalapino} stripes with hole density $x_s=1/2$ are
found in agreement with experiments \cite{Tranquada}.

Insight into pairing and stripe formation can been obtained by
adding or removing one, two or more particles/holes in the half-filled AF.  
When the hole moves a distance of $l$ lattice distances
the kinetic energy is reduced by $t/l^2$ but the AF is
frustrated by an energy $\sim lJ$.
Solving such a simple linear string-like 
model for the one-hole problem in a 2D AF for  $J\la t$
leads to a hole localized within a distance $l\sim (t/J)^{1/3}$ with energy 
\bea
   \varepsilon_1=-4t+\beta J^{2/3}t^{1/3} \,.
\eea
Within the WKB approximation one obtains $\beta=(3\pi/4)^{2/3}\simeq 1.8$.
Self consistent Born approximations \cite{Lee} find
similar hole energy $\varepsilon_1/t=-3.28+2.16(J/t)^{2/3}$ 
valid when $J\la0.4t$.

Because two nearby holes of opposite spin
can move through the AF ordered lattice without 
upsetting the
AF order they are not localized. If the pair has an energy that
is lower than that of two separate holes 
$\varepsilon_2>2\varepsilon_1-\varepsilon_0$, the odd-even difference can
as in nuclear physics be taken as an
effective pairing energy 
$\Delta =  (\varepsilon_0+\varepsilon_2)/2-\varepsilon_1$.

Adding more holes one might expect that more pairs form, eventually creating
a molecular Bose-Einstein condensate (mBEC) as in the BCS to BEC crossover in
3D ultracold atomic traps \cite{Leggett,Levin}. 
On the square lattice such
pairs have d-wave symmetry and mBEC will be a dSC 
\cite{Fradkin}. 
However, a number of numerical calculations
as well as experiments find that the holes form periodic stripes rather than forming a mBEC.
The equi-distance of the stripes indicate that they
``repel'' each other even in calculations without 
Coulomb frustration.
As a consequence the chemical potential must decrease with hole density or
equivalently $\varepsilon''\equiv (d^2\varepsilon/dx^2)_{x=0}>0$, and 
we can expand the stripe energy density  as
\bea \label{es}
   \varepsilon = -J-\varepsilon_1 x + \frac{1}{2}\varepsilon'' x^2 
                 + {\cal O}(x^3)\,,
\eea
for small doping. It is the second derivate $\varepsilon''$ of the
energy density at small doping that distinguishes the phases. It is
positive for the PM, FM and stripe phases, zero for PS but negative
for the unstable AF.

The stripe phase is a specific ordered mixed AF and PM phase and is a
continuous transition between the two pure phases as function of
density.  Similar mixed phase solutions are believed to occur in
neutron star crusts between nuclear matter and a neutron gas \cite{Ravenhall},
and possibly also between quark and nuclear matter \cite{QM}.
In both cases Coulomb energies add complexity to the mixed phases by ordering
them into structured crystallic phases.

\section{Coulomb frustration}

Long range Coulomb interactions prevent phase separation into two bulk
phases of different charge density. They also inhibit 
the formation of localized holes, pairs and stripes. The phase diagrams
of cuprates with such Coulomb frustration and optical lattices without
may therefore be very different.

Coulomb frustration in cuprates has been discussed in connection with
stripes (see e.g. \cite{Seibold,Fradkin}). Generally the CDW hole
pairs and stripes are energetically less favorable due to long range
Coulomb repulsion.  In the following we shall consider the stripes as
rods with charge less than the surrounding phase and calculate the
additional Coulomb energy of such structures.

Coulomb energies have been calculated for structures of various dimensionality
$D$ and volume filling fraction $f$ \cite{Ravenhall}
\bea \label{Ec}
   {\varepsilon}_C = \frac{2\pi}{D+2}\Delta\rho^2 R^2 
    f\left[ \frac{2}{D-2}(1-\frac{D}{2}f^{1-2/D})+f \right] \,.
\eea
Here, the charge density difference between the two phases is
$\Delta\rho = ex_s/a^3$ for the stripes and the volume filling fraction is  $f=x/x_s$.
The filling fraction is also the
inverse of the stripe distance in lattice units.
The dimensionality is $D=3$ for spherical droplets or bubbles,
$D=2$ for rods and tubes and $D=1$ for plate-like structures.
The diameter of the spheres, rods or the thickness of the plates is of order the
distance between layers $2R\sim  a\sim 4\AA$.
The stripes are rods in a 2D plane but are embedded in a 3D
layered structure.
For $D=2$ the expression in the square bracket of Eq. (\ref{Ec}) reduces
to $[\ln(1/f)-1+f]$. The logarithm originates from the Coulomb integral
$\int^ldz/z$ along the rod length $z$, which is cutoff by other rods at
a length scale $l\sim a\sqrt{f}$.
For the cuprates we furthermore reduce the Coulomb field by a dielectric
constant of order $\epsilon\sim5$.
The resulting Coulomb energy of stripe or rod-like structures $D=2$ is 
\bea \label{Ecc}
  \varepsilon_C \simeq \frac{\pi}{8}\frac{x_se^2}{a^4\epsilon} 
   f \left[\ln(1/f)-1+f\right] \,.
\eea

Energy costs associated with the interface structures are usually
added. Such surface energies are difficult to calculate for the stripes
because their extent is only a single lattice constant. In principle
they are already included in the stripe models. We will therefore
just add the Coulomb energies given above. However, the Coulomb energies and
the energy of the systems as a whole,
may be reduced by screening and hole hopping into the AF
whereby $R$ increases but $\Delta\rho$ and $f$ are reduced.

Inserting numbers $e^2/\hbar c=1/137$, $\epsilon=5$, $x_s=1/2$,
$a=4\AA$ we find that $\varepsilon_c\simeq 150$~meV$x[\ln(1/f)-1+f]$.
In comparison the energy gain by changing phase from an AF to a stripe
or PM phase increases with doping as
$(\varepsilon_{AF}-\varepsilon_{s/PM})\sim \tilde{J}x$.  The linear
coefficient is $\tilde{J}=(\varepsilon_1-3J/2)$ for the stripe phase
but considerably smaller for the PM phase $\tilde{J}\simeq
50-100$~meV according to Fig. 1. The Coulomb energy of Eq. (\ref{Ecc})
thus dominates at small doping due to the logarithmic singularity
and therefore the AF phase of density $n\la 1$ is preferred.

The AF phase is the ground state as long as the Coulomb energy of Eq.
(\ref{Ecc}) exceeds $\tilde{J}x$ corresponding to doping less than  
\bea
   x_{AF}\simeq x_s\exp\left[-\frac{8}{\pi}\frac{\tilde{J}a\epsilon}{x_se^2}-1
         \right] \,.
\eea
Inserting the above numbers and $\tilde{J}\simeq J$ 
we find $x_{AF}\simeq 0.1$ which is within range of
the observed $|x_{AF}|\simeq 0.03$ for hole doped and $x_{AF}\simeq 0.15$
for particle doped cuprates. In MF the particle-hole asymmetry arises from the
next-nearest neighbor hopping $t'\simeq -0.3t$ and leads to
an AF phase extending from half filling up to a particle doped density
$n>1$. \cite{Langmann}
Note that it is not taken into account that with increasing doping the AF density approaches
that of the PM, and eventually the charge difference $\Delta\rho$
between the AF and PM and the resulting Coulomb energy become
sufficiently small that stripes are favored.

In the above picture the incommensurate stripe phases at small doping
arise due to Coulomb frustration when $t'=0$. At larger doping the stripes approach
and will eventually affect each other. Experimentally the stripes
undergo a transition from an incommensurate to a commensurate phase
at $x\simeq 1/8$ corresponding to a stripe periodicity of
four lattice spacings.

\section{Superfluidity}

Since phase separation and stripe phases are sensitive to Coulomb frustration,
we can ask the related question: does superfluidity occur in
2D optical lattices with Fermi atom as in HTc cuprates or is Coulomb
frustration required? The answer depends on the mechanism behind HTc
which twenty years after its discovery is still not well understood.
Like the ground state phases discussed above, the various models
disagree about the origin of dSC and its dependence on the Hubbard parameters,
doping and the influence of stripes and Coulomb frustration.
In certain models \cite{Fradkin}, stripes are a prerequisite for dSC.
In the following a simple semi-phenomenological model will be discussed
based on similar ideas that pairing take place on stripes in an AF background.

Standard BCS superconductivity is s-wave and requires an attractive
pairing interaction. HTc superconductivity is d-wave and is believed
to arise in the correlated state of the Hubbard model with purely
repulsive forces.  The effective pairing must be a result of highly
collective effects since the two-body interaction is strongly
repulsive which inhibits onsite s-wave pairing but favors
next-neighbor d-wave pairing between opposite spins due to
super-exchange.

An instructive example where superfluidity arises is the 1D Hubbard
model in a staggered or AF background magnetic field.  At low magnetic
field it is a normal metal as the standard 1D Hubbard model
\cite{Lieb} whereas for strong field it becomes superfluid for
$x\le(2-\sqrt{2})$ \cite{Batista}.  The staggered magnetic field
leads to a linearly increasing string potential between holes of
opposite spin and constitutes a strong effective pairing potential.
However, the stripes observed in experiments have antiphase domain
walls. The staggered magnetic field is therefore of opposite sign on
the two sides of the stripe and no net staggering magnetic field
results.  No superconductivity is thus expected on the 1D stripes if
detached from the surrounding 2D AF phase.

Instead, effective pairing correlations in two-leg ladder models are found
transversely to the stripes \cite{Fradkin}. Two features of
superfluidity are important: that the electrons or atoms must pair and they must
condense. These two features limit the critical temperature for weak and strong
pairing respectively. In the weak limit the critical temperature is
exponentially suppressed as in standard BCS 
$T_c\sim \exp(-1/g)$ (for 2D see Eq. (\ref{TvH})),
where $g$ is the product of the pairing attraction and the level
density. In the strong limit it is argued \cite{Fradkin} that 
$T_c= \hbar^2 n_s(T=0)/4m^*$ where $n_s$ is the superfluid density
and $m*$ the effective mass of the particle. The superfluid density is
limited by the density of holes and thus vanishes with decreasing doping.

A simple model will now be constructed that incorporates these ideas
for the inhomogeneous stripe phase
in a semi-phenomenological way. The weakly interacting limit, where
the effective pairing and $T_c$ vanish, is approached when the AF
correlations dissolve and the PM phase appears.  To estimate the
pairing in this BCS limit we assume for simplicity that the pairing
take place between nearest neighbor holes and, since it is driven by
the AF correlations, that its effective strength $V$ is proportional
to the volume of the AF.  Specifically, we shall assume that it
vanishes when there is only one AF stripe between each half filled
stripe, i.e.
\bea \label{V}
    V=V_0(1-2x/x_s) .
\eea
Here, the effective pairing strength must exceed $V_0\ge 2t$ and $V_0\ga 7.35t$
in order that holes can form s-wave and d-wave two-body bound states respectively
\cite{Nozieres}. In contrast the onset of superconductivity occurs for
$V_0>2t$ and $V_0>0$ respectively.

In the strongly interacting limit at low hole density we ensure that
the superfluid density vanishes with hole density by explicitly multiplying
the density of states by the hole density. 

Generally, we assume the pairing takes place between pairs in the stripes, although pairs
can tunnel between stripes. The derivation of the gap equation for holes then
follows that of standard BCS on a lattice (see e.g. \cite{Micnas}).
The pairing is expected to be d-wave with a gap
$\Delta_{\bf q}=2\Delta[\cos(q_x)-\cos(q_y)]$
since s-wave is suppressed by on-site repulsion \cite{Micnas,Nozieres}. 
The resulting dSC gap equation for the critical temperature $T_c$ is
\bea
   \frac{1}{V} = \frac{x}{x_s}\frac{1}{N}\sum_{\bf q} 
     \left[\cos(q_x)-\cos(q_y)\right]^2
                  \frac{\tanh(\e/2T_c)}{2\e} \,.
\eea
Here, it should be noted that the prefactor $x/x_s$ ensures that the 
level density sums up to the correct hole density on the stripes.
The chemical potential is determined by the hole density in the stripes
\bea
    x_s = \frac{1}{N}\sum_{\bf q}\tanh(\e/2T) \,.
\eea
For simplicity we assume that the holes act as free electrons on the lattice
with dispersion $\e=-2t[\cos(q_x)+\cos(q_y)]-\mu$.

For large critical temperature $T_c\ga t$ the integral on the r.h.s.
of the gap equation is simple and we obtain
$T_c=Vx/4x_s=V_0(1-2x/x_s)x/4x_s$.  The pairing strength is, however,
not strong enough to satisfy that condition.  For small critical
temperature and $x_s\simeq 1$ ($\mu=0$) the singularities at $\e=0$
allows us to calculate the gap integral to leading logarithmic orders
\bea \label{GE}
   \frac{t}{V} = \frac{1}{\pi^2} \frac{x}{x_s} 
       \left[\ln^2(2t/T_c)+0.4\ln(2t/T_c) +1.78 \right] \,.
\eea 
Besides the usual BCS log originating 
from the energy denominator another
log arises from the logarithmic level density due to the van Hove
singularity in a 2D lattice at half filling.
The resulting critical temperature from Eq. (\ref{GE}) is
\bea \label{TvH}
   T_c = 2t\exp\left[-\sqrt{\pi^2tx_s/(xV) -1.74}+0.2\right] \,.
\eea
The filled stripe
condition $x_s\simeq 1$ is, however, only valid for dopings 
$1/4\la x\la 1/2$. For lower doping the filling is smaller, e.g. 
$x_s\simeq 0.5$ for $x\le1/8$.
The lower density of holes generally reduces $T_c$ \cite{Micnas}.

The stripe superfluidity differs from a superfluid of hole pairs. The
latter phase would be similar to the molecular Bose-Einstein
condensate (mBEC) discovered in ultracold atomic 3D traps (without
lattices) when a dilute gas of attractive fermionic atoms cross the
unitarity limit (infinite scattering length) and cross over from a BCS
to a mBEC \cite{Leggett}. 
The critical temperature for the superfluid mBEC is of the
order of the Fermi energy in 3D
$T_c=\hbar^2(n/2\zeta(3/2))^{2/3}\pi/m\simeq 0.218E_F$ in the mBEC and
slightly larger in the unitarity limit \cite{Thomas}.  In 2D, 
however, the critical temperature for a BEC vanishes and
can therefore not explain HTc.

\begin{figure}
\includegraphics[scale=0.5,angle=-90]{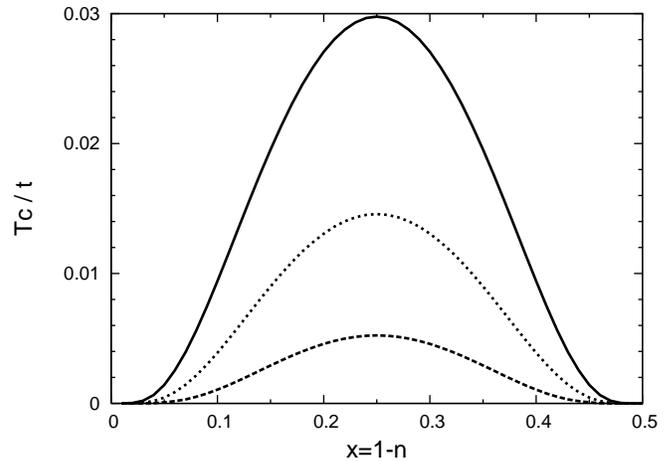}
\caption{Critical temperature vs. doping from Eq. (\ref{TvH}) for $V_0/t=2,3,4$
from bottom and up, and $x_s=1$.
\label{fig2}}
\end{figure}

In Fig. 2 the critical temperatures for dSC of Eq. (\ref{TvH}) is
shown for three effective pairing interactions. By construction d-wave
superfluidity vanishes near $x\sim 1/2$, where the effective pairing
$V$ vanishes, and also as $x\to 0$, where the density of holes
disappears. Thus the simple stripe pairing model gives a qualitative
description of HTc. In contrast d-wave superfluidity is largest at
low doping for standard uniform Hubbard models with an attractive
interaction and it extends to large doping. \cite{Micnas,Bruun}

As discussed above the semi-phenomenological stripe dSC model is just
one of many models for HTc dSC. It has the interesting consequence
that if PS occurs in optical lattices there will be no stripes and
therefore no dSC either. Both will have direct observational consequences
in optical lattices.

\section{Optical lattices in traps}

We can now address the phases present in optical lattices
based on the equation of states for the 2D Hubbard model
without Coulomb frustration, and calculate quantities measured
experimentally.

In the limit of many particles in a shallow confining potential, the
Thomas-Fermi approximation (TF) applies because the length scales
over which the trapping potential and density varies are long compared
to phase boundaries and the lattice spacing \cite{HH}. We shall in the
following assume that the confining potential is on a
harmonic oscillator form $V_2r^2$ as is the case in most experiments.
Within TF the total chemical potential is given by the sum of the trap
potential and the local chemical potential $\mu(n)=d\varepsilon/d n$,
where $\varepsilon$ and $n(r)$ are the local energy and number density
per site respectively, i.e.
\bea \label{mu}
  \mu_{Tot} = \mu(n) +V_2r^2 
      = \mu(n=0)+V_2R^2 \,.
\eea
It must be constant over the lattice and can therefore be set to its value
at the edge or radius $R$ of occupied lattice sites, which gives the last equation in
(\ref{mu}). The chemical potential for the dilute lattice gas in the 2D Hubbard
model is $\mu(n=0)=-4t$.

\begin{figure}
\includegraphics[scale=0.5,angle=-90]{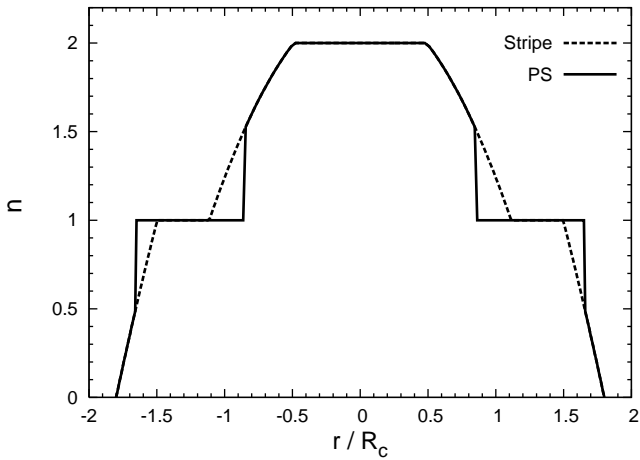}
\caption{Density distributions of Fermi atoms in an optical lattice
confined by a harmonic trap. $U=\av{\epsilon''}= 6t$ and $x_s=1/2$. 
Full (dashed) curves represents the PS (stripe) equation of states
(see text).
\label{fig3}}
\end{figure}

Measuring the density distribution $n(r)$ gives the 
chemical potential by inverting Eq. (\ref{mu}), and 
the equation of state can therefore be determined experimentally. 
We can from the above estimates for the equation
of states also predict the density distribution at least qualitatively.
The scale of the chemical potential is rather limited since it only varies
from $\mu(0)=-4t$ in the dilute limit to a value at half filling between
zero for the weakly repulsive PM ($U\ll t$) and $\mu(1)=4t$ for the
FM in the strongly repulsive limit ($U\gg t$).
Near half filling the chemical potential is $\mu(n=1)=\epsilon_1$ for
the stripe model. For the PS model the chemical potential is smaller
and constant between $0<x<x_s$, yet on a similar scale $\mu(n=1)\sim 2t$.
The average slope of the chemical potential is therefore
$\av{\epsilon''}=\mu(n=1)-\mu(n=0)\sim 6t$.
Approximating the chemical potential with such a linear density
dependence $\mu=\av{\epsilon''}n$ we obtain for $n\le1$ 
\bea \label{rho}
   n(r) = \frac{V_2}{\av{\varepsilon''}} (R^2-r^2) \,.
\eea 
The PM and stripe phases below half filling therefore extends up to a radius
$R_c=\sqrt{\varepsilon''/V_2}$.
The chemical potential $\mu(n)$ has a gap at half filling which is of order
$U$. As result a Mott insulator or AF 
density plateau appears at $n=1$, which extends a radial distance of order
$\sim \sqrt{U/V_2}$. The phases repeat for densities above half filling up
to the band insulator at $n=2$ such that the characteristic ``wedding cake layers''
appear as shown in Fig. 3.
The stripe and PS models differ qualitatively at densities around half filling
where the latter has a discontinuity at $n=n_s$ and $n=2-n_s$, where
we expect $n_s\simeq 0.5$ as discussed above.
The stripe phase is a d-wave superfluid 
between the AF and PM phases whereas no such intermediate
superfluid appears for the PS model. 
In both cases the AF phase is a MI at half filling $n=1$ only, whereas 
in the cuprates the AF extends between
densities $0.95-1.2$ due to Coulomb frustration or possibly a finite $t'$.
Detailed measurements of the densities in confined optical lattices near
half filling could therefore in principle reveal the effects of Coulomb frustration.
 
The density and radius are related through (\ref{mu}) and
the corresponding total number of atoms in the confined lattice is
\bea
   N=\frac{2\pi}{a^2} \int^R_0 n(r)rdr \,.
\eea
The critical numbers for the stripe and AF/MI phases to appear in the center
of the trap are
$N=\pi R_c^2/4a^2$ and $N=\pi R_c^2/2a^2$ respectively. Fillings
above half filling $n=1$ and up to the band insulator BI $n=2$ can be
reached by adding more atoms as shown in Fig. 3.  The critical
fillings and radii depend on the ratio of $U/\av{\epsilon''}$.
For $U\gg \av{\epsilon''}$ the MI plateau dominates the density distribution
and the number of particles required to reach higher densities and the
BI in the center is $N=\pi U/V_2a^2$.

Other possible scenarios are qualitative
different. If HTc is described by the 2D Hubbard model without Coulomb
frustration, then we can expect that the MI is replaced by an AF phase
around $n=1$ surrounded by two dSC phases at densities lower
(hole doped) and higher (particle doped). According to \cite{Becca} phases
with AF and dSC order are mixed when $U\la8t$ but coexist when
$U\ga8t$ with a first order transition between two densities.  A
critical point is predicted where a mixed AF+dSC phase terminate and
is replaced by a first order phase transition. Here the AFM and dSC
phases are separated and coexist with a density discontinuity. In
\cite{Aichhorn}, however, the pure dSC undergo a first order
transition to a mixed AF+dSC phase.  The densities at which these
transitions take place correspond to doping $x\simeq\pm 0.1$.  If
nearest neighbor interactions or next-nearest neighbor hopping are
included, the particle-hole symmetry is broken and the phase diagram
becomes asymmetric around half filling ($n=1$).
The models differ dramatically in their predictions of the 
phases in the shells surrounding the MI shell.

Experimentally the density distributions and shell structures
have been measured for Bose atoms in optical lattices
by a slicing method in combination with microwave transitions that differentiates
between singly and doubly occupied sites. \cite{Folling}
Very low temperatures and fine resolution will be required in order to observe the
density discontinuities of Fig. (3).

The momentum distributions of atoms in the confined lattice can be found
by time of flight experiments. After the system has
expanded a time $t$ the columb integrated densities are measured by
light absorption imaging.  In such time of flight experiments the
position is related to the momentum in the lattice before free
expansion as ${\bf r}={\bf k}t/m$ in the far field approximation.  As
a result one measures the momentum distribution $\langle n_{\bf
  k}\rangle$ and correlations $\langle n_{\bf k}n_{\bf k'}\rangle$ of
atoms in confined lattice.  In a sudden release $\langle n_{\bf
  k}\rangle$ is dominated by the Fourier transform of the tight
binding Wannier functions in the case of Fermi atoms whereas a BEC
displays characteristic Bragg peaks.  However, by adiabatically
turning off the optical lattice potential with Fermi atoms only the crystal momenta in
the lowest Bloch band remains, which for a 2D band insulator is a
square $|k_{x,y}|\le\pi$, as observed in \cite{Kohl}.  If the adiabatic
release can be made perfect at very low temperatures it is in
principle possible to measure accurate momentum distributions. In a
AF+PM mixed phase one should observe a two-component momentum
distribution: a square from the MI and a circular on top from the
PM. This also requires on-site repulsive interaction which must be
turned off suddenly or corrected for in the expansion.
  
The density-density correlation function is 
\bea
\langle \hat{n}_{\bf k}\hat{n}_{\bf k'}\rangle = 
\delta_{\bf k k'}\langle \hat{n}_{\bf k} \rangle + 
\langle \hat{a}_{\bf k}\hat{a}_{\bf k'}\hat{a}^\dagger_{\bf k'}\hat{a}^\dagger_{\bf k}\rangle
\,,
\eea
where the first term is the auto-correlations which is suppressed by a factor $1/N$ 
and can therefore be ignored. 
We proceed by expanding the field operator in terms of Wannier functions: 
$\hat{a}({\bf k})=w({\bf k})\sum_{\bf R}e^{-i{\bf kR}}\hat{a}({\bf R})$,
over the Bravais lattice ${\bf R}$.
We assume factorization into one-particle density matrices 
$\langle\hat{a}({\bf R})\hat{a}({\bf R'})\rangle=n_{\bf R}\delta_{\bf RR'}$
\cite{Bloch} which, however, 
excludes off-diagonal orders from phase coherence in a superfluid. Superfluids will
be discussed later.
In non-superfluid systems we then obtain the correlation function
\bea \label{corr}
 C({\bf k},{\bf k}') &=& \frac{
 \langle \hat{n}_{\bf k}\hat{n}_{\bf k'}\rangle }
 {\langle \hat{n}_{\bf k}\rangle\langle \hat{n}_{\bf k'} \rangle }
 \nonumber\\ &=&
 1 - \frac{1}{N^{2}}\sum_\sigma \left|\sum_{\bf R} e^{i({\bf k-k'}){\bf R}} n_{{\bf R}\sigma} 
 \right|^2 \,,
\eea
with $\langle \hat{n}_{\bf k}\rangle=N|w({\bf k})|^2$.
For bosons the minus sign is replaced by a positive one and gives the
characteristic Hanbury-Brown \& Twiss bunching observed for photons in
stellar interferometry and for mesons in relativistic heavy ion
collisions \cite{Baym}.  For non-interacting Fermi atoms the minus sign simply
enforces the Pauli principle which inhibits two atoms occupying the
same momentum state.

For MI and BI phases of bosons and fermions the occupation
numbers $n_{\bf R}$ are numbers, and the sum in Eq. (\ref{corr}) yields
Bragg peaks and dips respectively whenever the momentum difference
equals the reciprocal lattice vectors, i.e. ${\bf q}={\bf k}-{\bf
  k}'=2\pi(n_x,n_y)$, where $n_{x,y}$ are integers and in units $a=1$. The radius of
the occupied sites in the optical lattice limits the sum in
Eq. (\ref{corr}) and results in a finite momentum width of the peaks
and dips of order the inverse radius.

Bragg peaks have been observed for bosons in 3D \cite{Folling} and 2D
\cite{Spielman} lattices, and dips for 3D fermions in \cite{Rom}.  The
Bragg peaks and dips occur in $C({\bf k},{\bf k}')$ when ${\bf
  q}=\pi(n_x,n_y)$, where $n_x,n_y$ are even integers.  In an AF phase
the periodicity of a given spin is two lattice distances and
anti-bunching also appears for odd integers by an amount \cite{Bruun}
$C({\bf k},-{\bf k})=[(Um/2)/2E_{\bf k}]^2$, where $m$ is the AF order
parameter and $E_{\bf k}=\sqrt{(Um/2)^2+\ek^2}$ the quasiparticle
energy.

Just as for the AF phase we can in a stripe phase expect charge and
spin correlations as in low energy magnetic neutron scattering, i.e.
${\bf q}={\bf Q}_c$ and ${\bf q}={\bf Q}_s$
respectively, observed as
anti-bunching at these wave-numbers.  However, because the doping $x$
varies in the confined optical lattice, the Bragg dips are distributed
over the range of values for $x/x_s$ and are therefore hard to
distinguish from the background. If, however, the four stripe
periodicity with $x/x_s=1/8$ occurs for $1/8\le x\le 1/4$ as in HTc,
we can expect novel Bragg dips for charge correlations at ${\bf
  q}=(0,\pm\pi/2)$ and ${\bf q}=(\pm\pi/2,0)$ and for spin
correlations at ${\bf q}=\pi(1,1\pm 1/4)$ and ${\bf
  q}=\pi(1\pm1/4,1)$.

Pairing occurs between opposite momenta and leads to bunching for
${\bf k}=-{\bf k}'$.  Recently, phase coherence and s-wave pairing has
been observed in optical lattices with attractive onsite interactions
near the BCS-BEC cross-over \cite{Ketterle}.  As the lattice heights
are increased the MI phase dominates and phase coherence is gradually
lost. For repulsive on-site interactions the weaker d-wave
superfluidity may appear as discussed above.
We can expect atoms bunching by the amount \cite{Bruun}
$C({\bf k},-{\bf k})=\Delta_{\bf k}^2/2E_{\bf k}^2$, where the
quasiparticle energy is $E_{\bf k}=\sqrt{\Delta_{\bf k}^2+\ek^2}$.
If dSC is associated with stripes, the bunching due to d-wave superfluidity
should occur in conjunction with the stripe anti-bunching.

\section{Summary}

In conclusion, ultracold fermionic atoms in confined optical lattices
can reveal details of the equation of state and phases of the 2D
Hubbard model for a variety of interaction and hopping parameters.
Specifically two plausible models have been investigated with phase
separation and stripe formation respectively that results in
qualitatively different density distributions in optical lattices with
and without density discontinuities. However, in the cuprates Coulomb
frustration can induce stripe formation in both cases which makes it
hard to distinguish between the two cases. Coulomb frustration could
as shown in a simple model also be responsible for an
antiferromagnetic phases extending from densities between $0.95-1.15$
in cuprates whereas in optical lattices only $n=1$ MI plateau should
form because there is no Coulomb frustration. If stripes are required
for high temperature superconductivity, one will not observe
superfluidity in optical lattices if phase separation occurs.

We conclude that the various PM, FM, AF, d-wave superfluidity and
possibly also stripe phases can be observed by measurements of
densities, momentum distributions and correlations in expanding
ultracold atoms from optical lattices.  Bunching and anti-bunching
correlations can reveal these phases at characteristic Bragg momenta.
By varying the filling densities, interaction strengths and
temperatures, the onset and amount of the various phases can be studied
in order to determine the equation of states.  Experimental
determination of AF, stripe, metal and/or d-wave superfluid phases
will severely restrict the models, and the density and temperature
dependence on parameters such as $U,t,t'$ will provide much
understanding of the Hubbard model and HTc.

\vspace{-.4cm}
%\section*{References}


\begin{thebibliography}{99}
\vspace{-1.4cm}

\bibitem{Stoferle} T. St\"oferle {\it et al.}, Phys. Rev. Lett. {\bf 96}, 030401 (2006).  
\bibitem{Folling} S. F\"olling et al., Nature {\bf 434}, 481 (2005).
\bibitem{Ketterle} J.K. Chin, D. E. Miller, Y. Liu, C. Stan, W. Setiawan, C. Sanner, 
K. Xu, W. Ketterle, Nature {\bf 443}, 961 (2006).

\bibitem{Spielman} I.B. Spielman, W.D. Phillips, and J.V. Porto,
Phys. Rev. Lett. {\bf 98}, 080404 (2007).
%cond-mat/0606216.
\bibitem{Kohl} M. K\"ohl et al., Phys. Rev. Lett. {\bf 94}, 080403 (2005).
\bibitem{Rom} T. Rom, Th. Best, D. van Oosten, U. Schneider, S. Foelling, 
B. Paredes, I. Bloch, Nature {\bf 444}, 733 (2006).

\bibitem{Langmann} E. Langmann and M. Wallin, cond-mat/0406608.
\bibitem{Moreo} A. Moreo, D. Scalapino, and E. Dagotto, 
Phys. Rev. B  {\bf 43}, 11442 (1991).
\bibitem{Becca} F. Becca, M. Capone and S. Sorella, Phys. Rev. B  {\bf 62}, 12700 (2000).
\bibitem{Zitzler} R. Zitzler, Th. Pruschke, and R. Bulla, 
Eur. Phys. J. B  {\bf 27}, 473 (2002).
\bibitem{Aichhorn} M. Aichhorn and E. Arrigoni, Europhys. Lett. {\bf 71}, 117 (2005).
\bibitem{Macridin} A. Macridin, M. Jarrell and Th. Maier, 
Phys. Rev. B {\bf 74}, 085104 (2006). 
%cond-mat/0506148.
\bibitem{Zaanen} J. Zaanen and O. Gunnarsson, Phys. Rev. B {\bf 40}, 7391 (1989).
D. Poilblanc and T.M. Rice, Phys. Rev. B {\bf 39}, 9749 (1989). H.J. Schulz, 
J. Physique, 50, 2833 (1989); K. Machida, Physica C {\bf 158}, 192 (1989).
\bibitem{Scalapino} D.J. Scalapino, Physics Reports {\bf 250}, 329 (1995);
and cond-mat/0610710, to appear as Chapter 13 
in "Handbook of High Temperature Superconductivity", 
J. R. Schrieffer, editor, Springer, 2006.
S.R. White and D.J. Scalapino, Phys. Rev. Lett. {\bf 81}, 3227 (1998).
\bibitem{Tranquada} J.M. Tranquada et al., Phys. Rev. B {\bf 54}, 7489 (1996).
\bibitem{Sorella} S. Sorella et al., Phys. Rev. Lett. {\bf 88}, 117002 (2002).
\bibitem{Capone} M. Capone and G. Kotliar, Phys. Rev. B  {\bf 74}, 054513 (2006).

\bibitem{CC} M. Aichhorn, E. Arrigoni, M. Potthoff, and W. Hanke,
Phys. Rev. B {\bf 74}, 024508 (2006).

\bibitem{Hofstetter} W. Hofstetter {\it et al.},
Phys. Rev. Lett. {\bf 89}, 220407 (2002).

\bibitem{Kivelson} V.J. Emery and S.A. Kivelson, Physica C {\bf 209}, 597 (1993).
\bibitem{Hubbard} J. Hubbard, Phys. Rev. B {\bf 17}, 494 (1978).
\bibitem{Micnas} R. Micnas, J. Ranninger, S. Robaszkiewicz and S. Tabor,
 Phys. Rev. B {\bf 37}, 9410 (1988).
\bibitem{Kusko} C. Kusko and R.S. Markiewicz, cond-mat/0102440.
\bibitem{Aima} T. Aima and M. Imada, J. Phys. Soc. Jpn. {\bf 76}, 113708 (2007).
\bibitem{Fradkin} E. Arrigoni, E. Fradkin and S.A. Kivelson,
Phys. Rev. B {\bf 69}, 214519 (2004). 
%cond-mat/0309572.
\bibitem{Leggett} A.J. Leggett, in 
{\it Modern Trends in the Theory of Condensed Matter}, 
ed. A. Pekalski and R. Przystawa, Lect. Notes in Physics Vol. {\bf 115}
(Springer-Verlag, 1980), p. 13.
\bibitem{Levin} C. Chien, Y. He, Q. Chen and K. Levin, cond-mat/07063417. 

\bibitem{Lee} P.A. Lee, N. Nagaosa, and X-G. Wen,
Rev. Mod. Phys. {\bf 78}, 17 (2006).  
%cond-mat/0410445, 
\bibitem{Ravenhall} C.P. Lorenz, D.G. Ravenhall and C.J. Pethick,
 Phys. Rev. Lett. {\bf 70}, 379 (1993)
\bibitem{QM} H. Heiselberg, C. J. Pethick and E. F. Staubo, Phys. Rev. Lett. {\bf 70}, 1355 (1993).
\bibitem{Seibold} G. Seibold, C. Castellani, C. Di Castro, and M. Grilli, cond-mat/9803184
J. Lorenzana, C. Castellani, C. Di Castro, Phys. Rev. B {\bf 64}, 235128 (2001). 

\bibitem{Lieb} E.H. Lieb and F.Y. Wu, Phys. Rev. Lett. {\bf 20}, 1445 (1968).
\bibitem{Batista} C.D. Batista and G. Ortiz,
Phys. Rev. Lett. {\bf 85}, 4755 (2000). 
%cond-mat/0003158
\bibitem{Nozieres} F. Pistolesi and Ph. Nozieres, 
 Phys. Rev. B {\bf 66}, 054501 (2002).
\bibitem{Thomas} J. Kinast et al., Science {\bf 307}, 1296 (2005). 
\bibitem{HH} H. Heiselberg, Phys. Rev. A {\bf 74}, 033608 (2006).

\bibitem{Bruun} B.M. Andersen and G.M. Bruun, Phys. Rev. A {\bf 76}, 041602(R) (2007) 
%cond-mat/0706.361.
\bibitem{Altman} E. Altman, E. Demler, and M.D. Lukin, Phys. Rev. A {\bf 70}, 013603 (2004)
\bibitem{Bloch} I. Bloch, J. Dalibard, and W. Zwerger, cond-mat/0704.3011.
To appear in Rev. Mod. Phys.

\bibitem{Baym} G.A. Baym, Act. Phys. Pol. B {\bf 29}, 1839 (1998). 

\end{thebibliography}
\end{document}